\documentstyle[twocolumn,aps]{revtex}
\input{epsf.tex}

\def\be{\begin{equation}}
\def\ee{\end{equation}}
\def\ba{\begin{array}}
\def\ea{\end{array}}
\def\bea{\begin{eqnarray}}
\def\eea{\end{eqnarray}}
\begin{document}
\draft
\title{\bf Oriented collisions for cold synthesis of superheavy nuclei}
\author{Raj K. Gupta, M. Balasubramaniam, Rajesh Kumar and Narinder Singh\\
{\it Physics Department, Panjab University, Chandigarh-160014, India.}\\
}
\date{\today}
\maketitle

\begin{abstract}
The conditions of optimum orientations (lowest barrier and largest interaction radius) for 
deformed colliding nuclei are introduced in "cold" fusion of superheavy nuclei. Also, the 
role of (octupole and) hexadecupole deformations is studied. We have used the proximity
potential and applied our method to Ca-induced reactions.
\end{abstract}

PACS number(s): 25.60.Pj, 25.70.Gh, 27.90.+b \\

Collisions between the deformed, oriented nuclei have been of much interest from time to time.
In early 1980's it got trigged off, beginning with a suggestion by Greiner \cite{greiner81} 
that oriented $^{238}$U+$^{238}$U collisions could lead to a very long lived (life-time 
$\sim {10}^{-20}$ sec) giant molecule. A number of calculations were made 
\cite{baltz82,muenchow82,r-brown83,seiwert84,malhotra85} which all resulted in showing that 
the barrier is lowered due to deformations and orientations of colliding nuclei and that it 
is lowest for the 0$^0$-180$^0$ orientations of two $^{238}$U nuclei, known as pole-to-pole 
(p-p) or nose-to-nose configuration. Note that $^{238}$U is a prolate deformed nucleus, and 
hence the above result is only for prolate-prolate collisions, though quoted in the literature 
loosely for all oriented collisions. In fact, we show in the present paper that colliding 
nuclei with different signs of their quadrupole deformations result in lowest barrier for 
different orientations (see Table 1). For example, for prolate-oblate collisions, the barrier 
is lowest for 0$^0$-90$^0$ (equator-equator crossed, in short, e-c) configuration, as is 
envisaged very recently by N\"orenberg \cite{noerenberg94}. Also, till recently \cite{misicu02}, 
the role of multipoles higher than the quadrupole had not been investigated. We find that the 
inclusion of higher multipole deformations is favorable for fusion in some cases only i.e. the 
barriers are lowered only for some orientations. Some of these results require immediate 
attention and verification by alternative methods.

We use here the quantum mechanical fragmentation theory (QMFT), extended to include the higher
multipole deformations and orientations degrees of freedom. In QMFT, cold 
synthesis of new and superheavy nuclei was first proposed by one of us and collaborators 
\cite{fink74,sandu76,gupta77}, where a method was given for selecting out an {\it optimum} 
"cold" target-projectile (T-P) combination. Cold compound systems were considered to be formed 
for {\it all} those T-P combinations that lie at the bottom of the {\it minima} in the 
potential energy surface of a given compound nucleus, calculated for all possible T-P
combinations, referred to as "cold reaction valleys" or reaction partners leading to "cold 
fusion" \cite{sandu76,gupta77,gupta77a,gupta77b,gupta77c}. This information on "cold fusion 
valleys" was further optimized \cite{gupta77} by the requirements of smallest interaction 
barrier, largest interaction radius and non-necked (no saddle) nuclear shapes, identifying the 
cases of "cold", "warm/ tepid" and "hot" fusion reactions. The key result behind the cold 
fusion reaction valleys is the {\it shell closure effects} of one or both the reaction partners. 
The QMFT was advanced as a unified approach both for heavy ion collisions, leading to fusion, 
and fission of nuclei including the cluster radioactivity (see e.g. the reviews in 
\cite{gupta99} and the references therein). 

We choose to apply our method to the recent experiments of highly neutron-rich $^{48}Ca$ beam 
bombarded on neutron-rich actinides $^{232}Th$, $^{238}U$, $^{242,244}Pu$ and $^{248}Cm$, 
forming the compound systems $^{280}110^*$, $^{286}112^*$, $^{290,292}114^*$ and $^{296}116^*$ 
\cite{oganess99}. In these reactions, for near the Coulomb barrier energies, the compound 
nucleus excitation energy $E^*\sim$30-35 MeV, in between the one for cold (10-20 MeV) and hot 
(40-50 MeV) fusion reactions. The use of neutron-rich (radioactive) nuclei is essential for overshooting the 
{\it centre} of island of superheavy nuclei (the next doubly magic nucleus) and their 
deformations and orientations could provide an added advantage since the fusion barrier gets 
lowered, or, in other words, the excitation energy of compound system gets reduced. This means 
a possibility that the "warm" and/or "hot" fusion reactions could also be reached in "cold" 
fusion, as is found to be the case here in the following calculations.

The QMFT is worked out in terms of the mass (and charge) asymmetry 
$\eta$=(A$_1$-A$_2$)/(A$_1$+A$_2$) (and $\eta_Z$=(Z$_1$-Z$_2$)/(Z$_1$+Z$_2$)), the relative 
separation $\vec R$, the deformations $\beta_{{\lambda}i}$, (so far $\lambda$=2 only, the 
quadrupole deformations) of two nuclei (i=1,2) or, in general, the two fragments, and the neck 
parameter $\epsilon$ \cite{gupta75,maruhn74,gupta80,yamaji77}. We introduce here the higher 
multipole deformations $\lambda$=3 and 4, i.e. the octupole and hexadecupole deformations, as 
additional new parameters. Also, two orientation angles $\theta_i$ are included, as in 
\cite{malhotra85} (see Fig. 1, illustrated for quadrupole deformations). So far, the 
time-dependent Schr\"odinger equation in $\eta$ is solved for non-oriented collisions and for 
weakly coupled $\eta$ and $\eta_Z$ motions:
\begin{equation} 
\label{eq:1}
H\Psi(\eta,t)=i\hbar \frac{\partial }{\partial t} \Psi(\eta,t),
\end{equation}
with R(t) treated classically, and $\beta_{2i}$ and $\epsilon$ fixed by minimizing the 
collective potential V(R,$\eta$,$\eta_Z$,$\beta_{2i}$,$\epsilon$). Eq. (\ref{eq:1}), solved 
for a number of heavy systems \cite{gupta80,yamaji77}, shows that a few nucleon to a large 
mass transfer occurs for T-P combinations coming from {\it outside} the potential energy minima,  
whereas the same is {\it zero} for a T-P referring to potential energy minima. This means that 
{\it for cold reaction partners}, the two nuclei stick together and form a deformed compound 
system. A few nucleon transfer may, however, occur if a "conditional" saddle exists 
\cite{malhotra86}. The solution of Eq. (\ref{eq:1}) is very much computer-time 
consuming, and hence the following (next paragraph) simplifications are exercised on the basis 
of calculated quantities.

The potentials V(R,$\eta$) and V(R,$\eta_Z$) for non-oriented nuclei, calculated
within the Strutinsky method by using asymmetric two-center shell model (ATCSM), show that 
the motions in both $\eta$ and $\eta_Z$ are much faster than the R-motion. This means that 
these potentials are nearly independent of the R-coordinate and hence R could be taken as a 
time-independent parameter. This reduces Eq. (\ref{eq:1}) to the stationary Schr\"odinger 
equation in $\eta$,
\begin{equation} 
\label{eq:2}
\{ - \frac{\hbar^2}{2\sqrt{ B_{\eta\eta} }} \frac{\partial}{\partial\eta}
\frac{1}{\sqrt{ B\eta\eta }} \frac{\partial}{\partial\eta} +
V(\eta)\} \Psi^{\nu}(\eta)=E^{\nu}\Psi^{\nu}(\eta).
\end{equation}
Here R is fixed at the post saddle point, a choice justified by many calculations 
\cite{gupta99}, and by an explicit, analytical solution of 
time-dependent Schr\"odinger equation in $\eta_Z$ coordinate \cite{saroha83}. An interesting 
result of these calculations is that the yields 
($\propto \mid\Psi$($\eta$)$\mid^2$ or $\mid\Psi$($\eta_Z$)$\mid^2$, respectively, for mass or 
charge distributions) are nearly insensitive to the detailed structure of the kinetic energy 
term in the Hamiltonian which consisted of Cranking masses B$_{\eta\eta}$ consistently 
calculated by using ATCSM. In other words, the static potential V($\eta$) or V($\eta_Z$) 
contains all the important information of a colliding or fissioning system. 

Since the potential V($\eta$,R) is nearly independent of the choice of R-value, for oriented 
nuclei, we define it as sum of two binding energies, and the deformation and orientation 
dependent Coulomb and proximity potentials: 
\bea
\label{eq:3}
V(\eta ,R)&=&-\sum_{i=1}^{2} B_i(A_{i},Z_{i},\beta _{{\lambda}i})+ 
             E_c(Z_{i},\beta _{{\lambda}i},\theta_i) \nonumber  \\
           &+&V_{P}(A_{i},\beta _{{\lambda}i}, \theta_i).
\eea
Here, the binding energies $B_i$ are taken from M\"oller et al. \cite{moeller95} 
for Z$\ge$8, and from experiments \cite{audi95} for Z$\le$7. The Coulomb and proximity 
potentials, with higher multipole deformations included, are obtained by following the works of 
\cite{wong73} and \cite{malhotra85}, respectively. The Coulonb potential 
\bea
\label{eq:4}
E_c&=&{{Z_1Z_2e^2}\over {R}}+ {3Z_1Z_2e^2}\sum_{\lambda ,i=1,2}{{1\over {2\lambda +1}}
{{R_{0i}^{\lambda}}\over {R^{\lambda +1}}}}Y_{\lambda}^{(0)}(\alpha _i)  \nonumber  \\
&\cdot&\Big [\beta _{{\lambda}i}+{4\over 7}\beta _{{\lambda}i}^2Y_{\lambda}^{(0)}(\alpha_i)\delta_{\lambda ,2}\Big ],
\eea
with 
\be
\label{eq:5}
R_i(\alpha _i)=R_{0i}\big [1+\sum _{\lambda}\beta _{{\lambda}i}Y_{\lambda}^{(0)}(\alpha _i)\big ],
\ee 
where $R_{0i}=1.28A_i^{1/3}-0.76+0.8A_i^{-1/3}$. A similar expression is obtained by Rumin et al. 
\cite{rumin03} which differs in the quadrupole interaction term proportional to $\beta _{2i}^2$.

The nuclear proximity potential
\be
V_P=4\pi\bar{R}\gamma b\Phi (s_0),
\label{eq:6}
\ee
where the specific nuclear surface tension coefficient 
$\gamma =0.9517\left[1-1.7826\left(\frac{N-Z}{A} \right)^{2}\right]$ MeV fm$^{-2}$;
the surface thickness $b$=0.99 fm; and the universal function, independent of the geometery 
of nuclear system, is
\be
\Phi (s_0)=\left \{
\ba{ll}
-{1 \over 2}(s_0-2.54)^2-0.0852(s_0-2.54)^3 \\
-3.437exp(-{s_0 \over 0.75}) 
\ea
\right.
\label{eq:7}
\ee
respectively, for $s_0\le 1.2511$ and $\ge 1.2511$, with
\be
\label{eq:8} 
s_0=[R-R_1(\alpha_1)cos\psi_1-R_2(\alpha_2)cos\psi_2]/b,
\ee
the separation distance between the colliding surfaces, parallel to R taken 
along the collision axis, in units of b. For $s_0$ to be minimum , i.e. 
$\partial s_0/\partial {\alpha_i}=0$, it follows from Fig. 1 that $\psi_1=\theta_1-\alpha_1$, 
$\psi_2=180-\theta_2-\alpha_2$ and $tan\psi_i=-R_i^{\prime}(\alpha_i)/R_i(\alpha_i)$. Finally, 
$\bar{R}$, the mean curvature radius characterizing the gap, for nuclei lying in the same 
plane, is ${{1/\bar{R}^2}}={{1/{R_{11}R_{12}}}}+{{1/{R_{21}R_{22}}}}
+{{1/{R_{11}R_{22}}}}+{{1/{R_{21}R_{12}}}}$,
where the four principal radii of curvature $R_{i1}$ and $R_{i2}$, at the points D (denoted 1)
and E (denoted 2) of minimum $s_0$, are given by Eq. (15) in Ref. \cite{malhotra85}. For further 
details, see \cite{malhotra85}. Recently, Misicu and Greiner \cite{misicu02} have also 
derived the heavy ion interaction potential by using a multipole expansion of the densities 
in a double folding proceedure. Such a proceedure is shown \cite{ismail02} to depend 
strongly on the number of terms included in the expansion. Three terms are found to be 
sufficient for the internal region of the nuclear potential, whereas up to five terms are shown 
necessary for the physically more relevant surface and tail region for heavy ion collisions.

For the fixed orientations, the charges Z$_i$ in (\ref{eq:3}) are fixed by minimizing the 
potential $V(R,\eta ,\eta _Z,\beta _{{\lambda}i},\theta_i)$ in $\eta_Z$ coordinate (which fixes 
the deformation coordinates $\beta_{{\lambda}i}$ also). In Eq. (\ref{eq:8}), for  fixed R, 
s$_0$ is different for different orientations, and for fixed s$_0$, R is different for 
different orientations which is used here in the following.

Table 1 gives the orientations of nuclei for the lowest barrier, for all possible combinations 
of different signs of their quadrupole deformations (prolate, oblate or spherical). These 
barriers also lie at the largest interaction radii. Fig. 2 illustrates our result for 
prolate-oblate $^{238}$Pu+$^{48}$Ar$\rightarrow ^{286}$112 reaction (see solid lines, 
where deformations are included to all orders, $\lambda$=2,3,4). According to the QMFT 
\cite{gupta77}, as already stated above, the above conditions are for an optimum cold fusion 
reaction. In other words, the orientaions in Table 1 are the optimum orientations for cold 
fusion reactions using deformed nuclei. We further notice from Fig. 2, that the inclusion of 
higher multipole deformations is not always favorable for fusion (compare solid lines, with 
dotted ones for $\lambda$=2 alone): the addition of $\beta_{4i}$ term ($\beta_{3i}$=0) lowers 
the barriers for some sets of oientations whereas it raises them for the other sets of 
orientations (illustrated in Fig. 2 for two cases each). Thus, the choice of nuclei having 
octupole and hexadecupole deformations for (cold) fusion reactions must be made judiciously, 
depending on not only the signs of their quadrupole deformations but also the orientation 
angles.

Fig. 3 shows the fragmentation potentials for optimum orientations of the different T-P 
combinations, at a fixed separation $s_0$=1.5 fm, forming the same compound nucleus 
$^{286}$112$^*$. The case of spherical nuclei \cite{kumar03} is also plotted for comparisons. 
Apparently, due to deformation and orientation degrees of freedom, all the potential energy 
minima are lowered, some new minima have appeared and some old ones have disappeared. 
Specifically, new deep minima occur at $^{56}$Cr+$^{230}$Ra and $^{106}$Mo+$^{180}$Yb, which are 
in addition to the ones referring to the region of cluster radioacticity and/ or "hot" fusion 
(involving light nuclei of masses $<$30). The minima that have disappeared refer to well known 
cases of Ca and Pb (or neighbouring) nuclei; here $^{48,50}$Ca+$^{238,236}$U and 
$^{80}$Ge+$^{206}$Hg combinations. Similar results are obtained for the compound systems 
$^{290,292}114^*$ and $^{296}116^*$. For $^{280}110^*$, however, Ca and Hg minima are still 
deep and could be used as cold fusion reactions (details to be published elesewhere). 
We further notice from Fig. 2 that, w.r.t. the g.s. energy, all minima now refer to much 
smaller excitation energies, some of them lying even below it. For example, for Ca minima it 
is reduced from $\sim$35 MeV for spherical nuclei to $<$20 MeV for deformed and oriented 
collisions. This means that, as compared to spherical nuclei, oriented collisions result in 
cooler fusion reactions. Furthermore, all the optimum cold oriented collisions involve 
radioactive nuclei.  

The above results are seen better in the calculated mass distribution yields
$Y(A_i)=\mid \Psi (\eta (A_i))\mid ^2 {\sqrt {B_{\eta \eta}}} {2\over A}$, for ${\nu =0}$.
Here, $\Psi ^{(\nu )}$ are the solutions of Eq. (\ref{eq:3}) and ${B_{\eta \eta}}$ are the 
classical hydrodynamical masses \cite{kroeger80}. We take the view that, since fragments 
related to the minimum in $V(\eta)$ are more probable, the yields must give the intermediate 
(two) fragment formation yields or, in short the formation yields for a cool compound nucleus 
\cite{kumar03}, where the contribution of barrier penetration is not included. Evidently, for
oriented collisions, the yields for new T-P pairs $^{56}$Cr+$^{230}$Ra and $^{106}$Mo+$^{180}$Yb
are larger than for their neighbouring Ca and Hg induced reactions. The (near) symmetric 
combination has the largest yield, but they are known to form necked-in shapes, signifying 
preformation of fission fragments \cite{gupta77,gupta99,kumar03}.  

Summarizing, we have extended the QMFT for use of oriented collisions and inclusion of higher
multipole deformations, which result in the reduction of excitation energies of the compound 
system formed due to different T-P combinations. This means that both the "warm" and "hot" 
fusion reactions could now be reached in "cold fusion" also. The idea of optimum orientations 
for cold fusion reactions is introduced for the first time, which leads to new cold fusion 
reaction partners. The choice of nuclei with hexadecupole deformations is shown to depend 
strongly on both the signs of their quadrupole deformations and orientation angles.
\begin{table}[ht]
{\bf Table 1:}
{The optimum orientations for "cold" fusion of nuclei with all possible combinations of 
deformations. Here the spherical nuclei (denoted by $\dagger$) are considered to have zero 
octupole and hexadecupole deformations.
}
\begin{center}
\begin{tabular}{| c| c|| c| c|}
Nuclear&Optimum&Nuclear&Optimum   \\ 
deformations&orientations&deformations&orientations \\ \hline
Prolate-Prolate&$0^0-180^0$&Prolate-Spherical&$0^0-\dagger$\\
Oblate-Oblate&$90^0-90^0$&Oblate-Spherical&$90^0-\dagger$\\
Prolate-Oblate&$0^0-90^0$&Spherical-Prolate&$\dagger-180^0$\\
Oblate-Prolate&$90^0-180^0$&Spherical-Oblate&$\dagger-90^0$\\  
\end{tabular}
\end{center}
\end{table}


\newpage
{\bf Figure Captions} \\
\begin{description}
\item {Fig. 1} {Schematic configuration of two axially symmetric deformed, oriented nuclei, 
lying in the same plane ($\phi=0^0$).}

\item {Fig. 2} {Scattering potentials for the prolate-oblate 
$^{238}$Pu+$^{48}$Ar$\rightarrow ^{286}$112$^*$, at different orientations. The R-values at 
the top of the three lowest lying barriers are also shown.}

\item {Fig. 3}{Fragmentation potentials of $^{286}$112$^*$ for the optimum orientations of different 
T-P combinations with $\lambda$=2,3,4 (solid line with symbols) and for spherical nuclei (solid 
line). For Z$\le$7, the $\beta _{2i}$ are from relativistic mean field calculations using TM2 
force \cite{mehta03}, and for Z$>$7 from \cite{moeller95}. The $\beta_{3i}=\beta_{4i}=0$ for 
Z$\le$7. For the spherical case, $\beta_{2i}=\beta_{3i}=\beta_{4i}=0$. The g.s. is the ground 
state energy.}

\item {Fig. 4}{Calculated yields of $^{286}$112 for optimum orientations of different 
T-P combinations, with $\lambda$=2,3,4 (solid line with symbols) and spherical nuclei 
(solid line). }

\end{description}
\vfill\eject

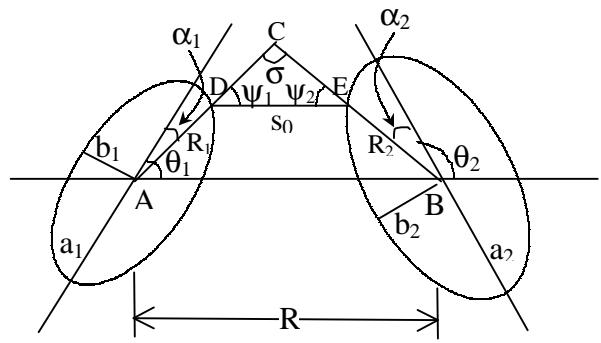
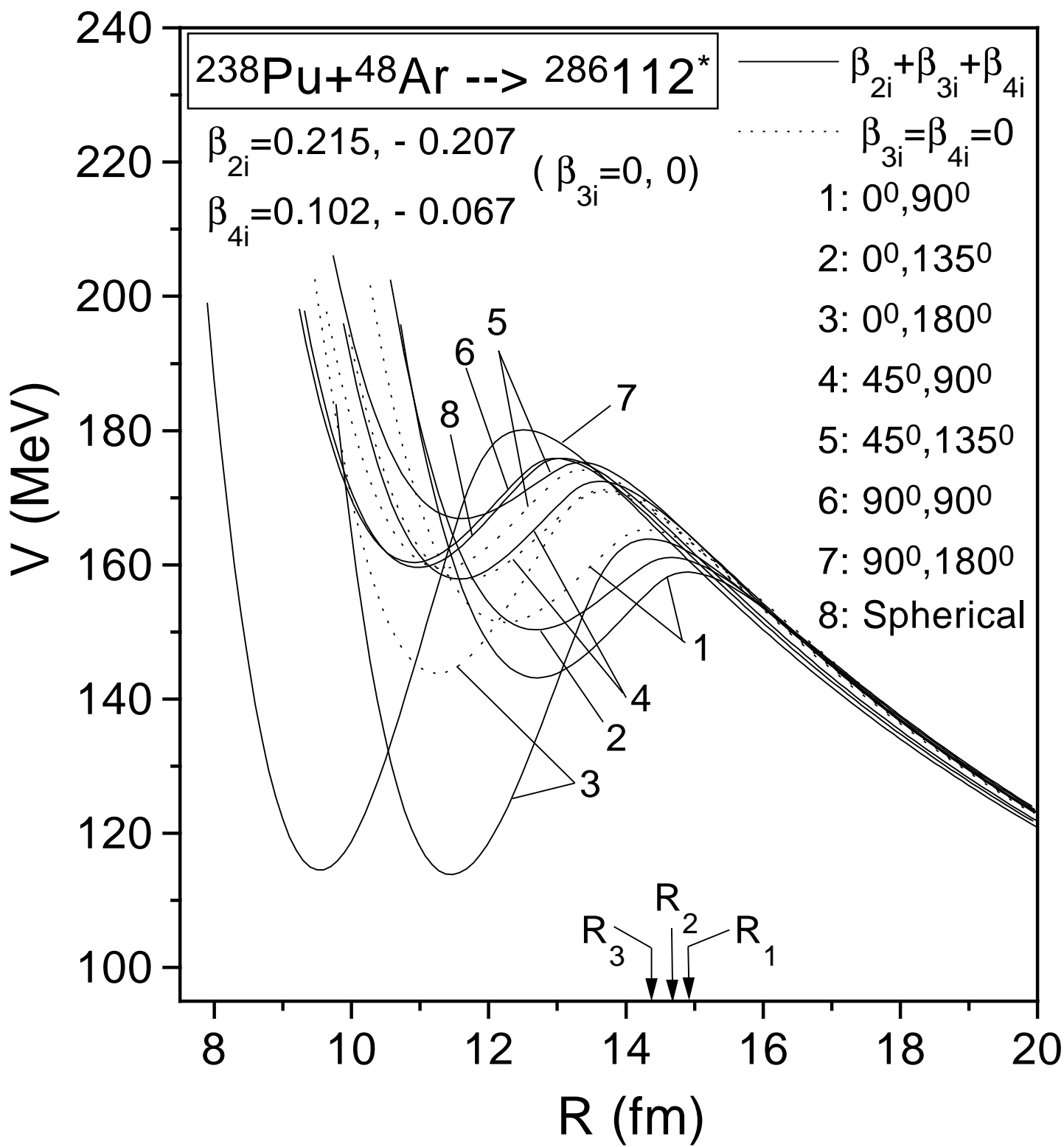
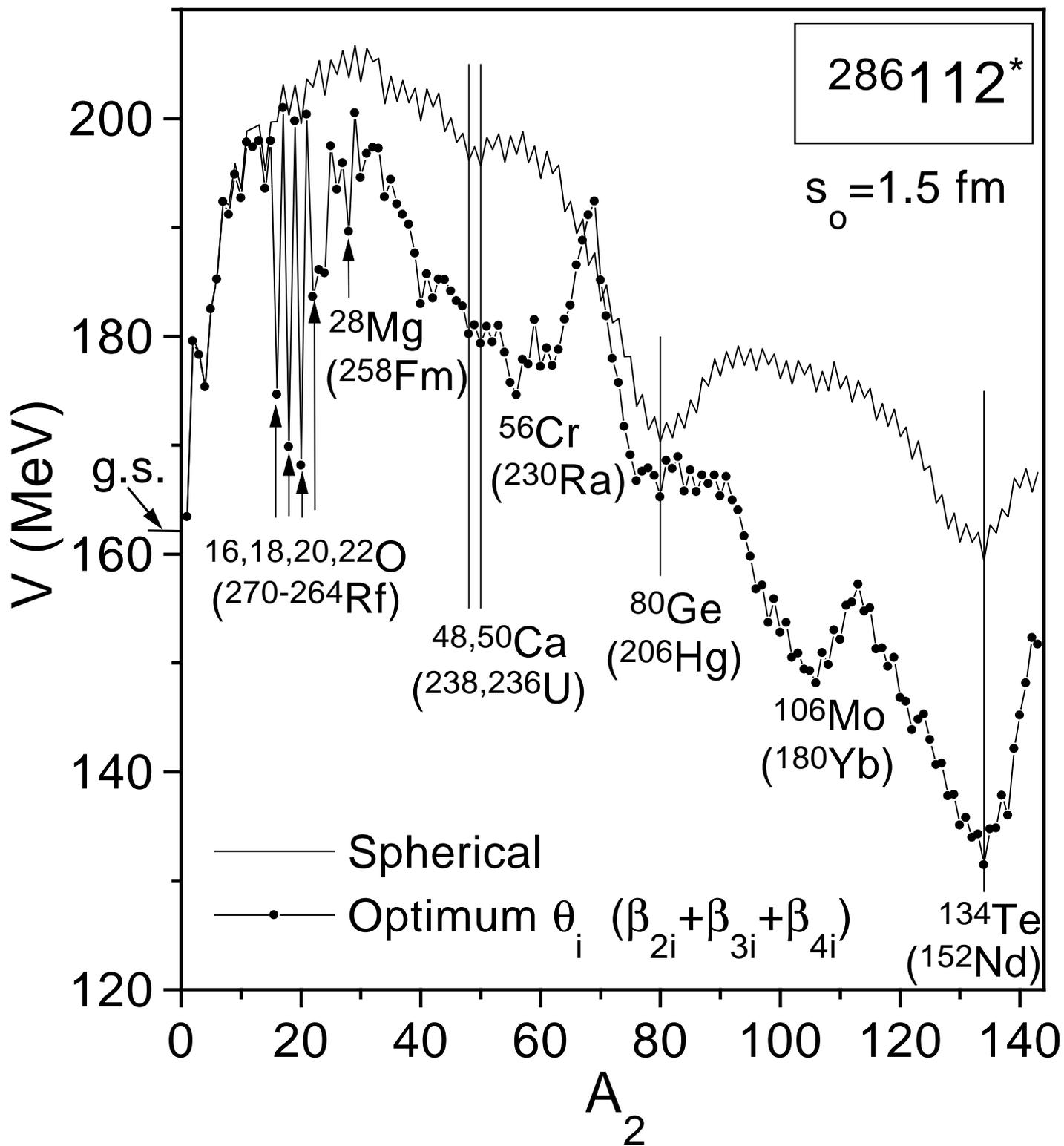
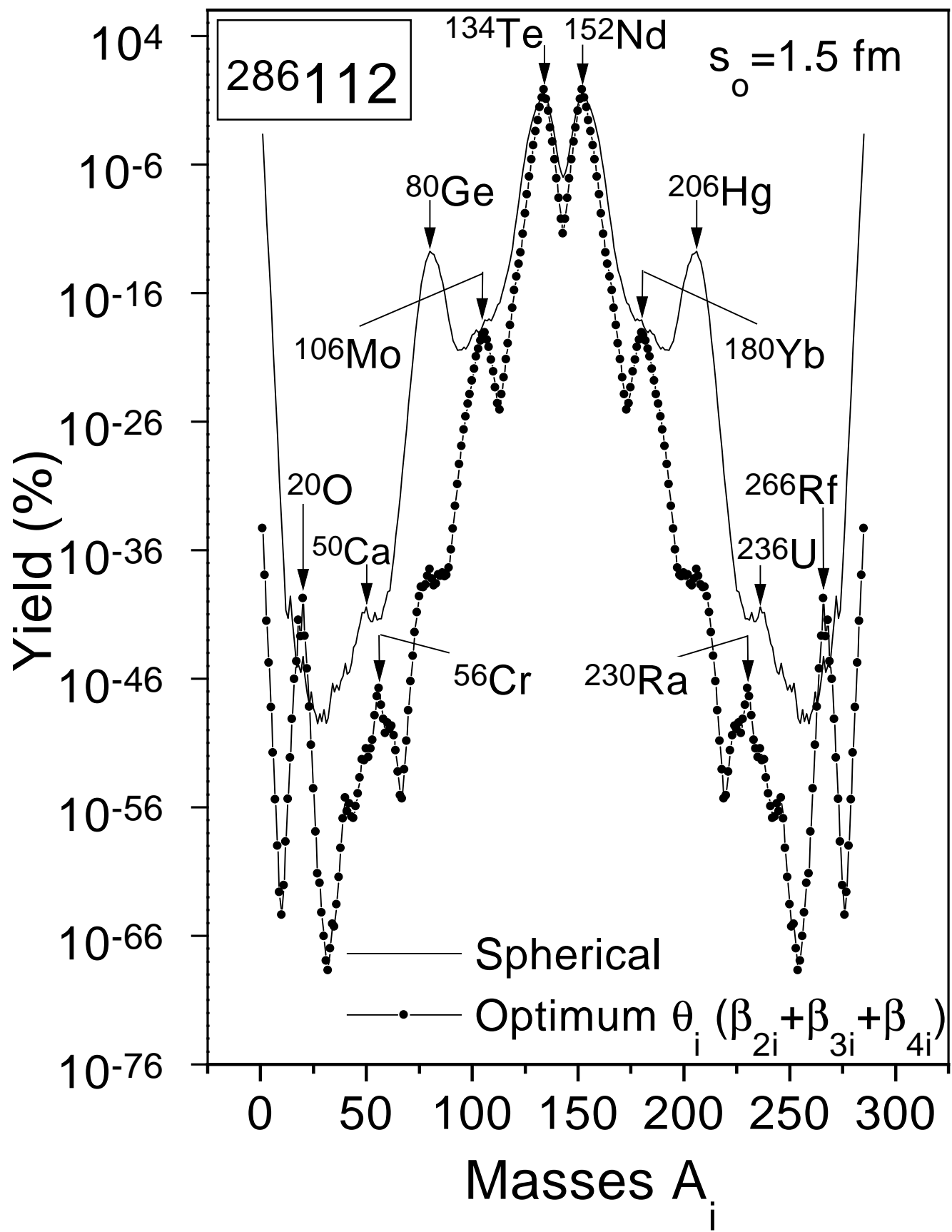
\end{document}